\documentclass{ptephy}

\begin{document}

\title{($d,^3$He) reaction on odd-neutron nuclear target for the formation of deeply bound pionic atoms}

\author{\name{Natsumi Ikeno}{1,\ast}, \name{Junko Yamagata-Sekihara}{2,3}
\name{Hideko Nagahiro}{1}, and
\name{Satoru Hirenzaki}{1}}

\address{\affil{1}{Department of Physics, Nara Women's University, Nara 630-8506, Japan}
\affil{2}{Research Center for Physics and Mathematics, Osaka Electro-Communication University, Neyagawa, Osaka 572-8530, Japan}
\affil{3}{Institute of Particle and Nuclear Studies, High Energy Accelerator Research Organization (KEK), 1-1, Oho, Ibaraki 305-0801, Japan}
\email{jan\_ikeno@cc.nara-wu.ac.jp}}

\begin{abstract}%
We consider the pionic atom spectroscopy by the ($d,^3$He) reaction 
on an odd-neutron nuclear target in this article, 
which has not been investigated so far.
In the ($d,^3$He) reaction on the odd-neutron nuclear target,
we can observe the pionic states in the even-neutron nucleus 
with spin-parity $0^+$. 
We expect that this pionic state does not have
the additional shifts
due to the effects of the residual interaction
between neutron-hole and pionic states.
For the even-neutron nuclear target cases, 
we may have to take into account the residual interaction effects 
to deduce the binding energies of the pionic states
precisely
from the high precision experimental data
since the final pionic states are the 
pion-particle plus neutron-hole
[$\pi \otimes n^{-1}$] states.
Thus, 
in addition to widening the domain of the pionic atom spectroscopy 
in nuclear chart,
the present study of the ($d,^3$He) reaction on the odd-neutron
target is considered to be
important to deduce extremely precise information on 
the binding energies of the observed pionic states, 
and to know the pion properties and aspects of the symmetries of the
strong interaction at finite density.

We modify the formula of the even-even nuclear target case to study 
the pionic atom formation spectra on the even-odd nuclear target
and show the numerical results of $^{117}$Sn target case.
This experiment will be performed at RIBF/RIKEN in near future.
\end{abstract}

\subjectindex{D15,D33,D25,D22}

\maketitle

\section{Introduction}

One of the most interesting subjects in the contemporary hadron-nuclear
physics is to study the aspects of the QCD symmetry in the extreme
conditions with high density and/or temperature~\cite{QGP}.
To know the hadron properties in these extreme conditions is an
important way to find the consequences of the change of the symmetry
breaking pattern of QCD 
based on the deep understanding of the origin of the hadron
properties such as their masses and interactions~\cite{QCD}.

Meson-nucleus systems are very interesting objects in this context to
deduce the hadron properties at finite density~\cite{Gal,Piatom}.
In Ref.~\cite{KSuzuki}, the partial restoration of chiral symmetry 
is concluded by determining the pion weak decay constant
$f_\pi$ from the observation of deeply bound pionic atoms using the
Tomozawa$-$Weinberg and Gell-Mann$-$Oakes$-$Renner relations.
The theoretical foundation of this scenario is strengthened by
Refs.~\cite{Kolo,Jido}.
The effects of $U_A(1)$ anomaly at finite density is also discussed in
Refs.~\cite{Nagahiro,Nagahiro2,Jido_etap,Nagahiro3,etap_Exp} with the possible existence 
and observation of the 
$\eta^\prime (958)$ mesic nucleus.
Kaonic~\cite{K,K2,K3,K4,K5} and $\eta$~\cite{eta,eta2,eta3,eta4,eta5}
mesic nucleus systems are also interesting because of
the existence of the resonances, $\Lambda(1405)$ and $N^\ast (1535)$, 
which are
strongly coupled to $\bar{K} N$ and $\eta N$ channels, respectively.

Within these meson-nucleus systems, 
the pionic atom is the best system
to perform accurate spectroscopic studies at present 
since we have the well-established theoretical and experimental techniques.
So far, based on the theoretical predictions~\cite{Toki,nd,nd_d3He,d3He,Umemoto2}, 
there exist in the world
four sets of the successful experimental results 
for the formation of the deeply bound pionic atoms in the ($d,^3$He)
reaction~\cite{Yamazaki,Gilg,Itahashi,Geissel,KSuzuki,Itoh}.
However, all theoretical and experimental efforts concentrated on the
even-even nuclear target cases.

In this article, we report a theoretical study of the deeply bound
pionic atom formation in the ($d,^3$He) reaction 
on the odd-neutron (odd-N) nuclear
target $^{117}$Sn.
The reasons why we consider the odd-N nuclear target cases are followings.

In the ($d,^3$He) reaction for the pionic atom formation 
on an even-neutron nuclear target, 
one neutron in the target nucleus is
picked-up and removed from the nucleus.
Consequently, there exists a neutron-hole state $j_n^{-1}$ in the
daughter nucleus which couples to pionic atom states ($n_\pi \ell_\pi$) to make
total spin $J$ state $[j_n^{-1} \otimes \ell_\pi ]_J$.
In these states, we generally expect the existence of the residual
interaction effects and of the level splittings between different $J$
states~\cite{Kaneyasu,Nose}.
The experimental errors of the pion binding energies and widths
reported in Ref.~\cite{KSuzuki} are 
approaching the magnitude of
the estimated level shifts and width changes due to the residual
interaction~\cite{Kaneyasu,Nose}.
Thus, the formation of the pionic states on the even-even nucleus by 
the ($d,^3$He) reactions on the odd-N nuclear target 
is preferable to avoid the additional difficulties due to the
residual interaction effects 
to determine the pion binding energies precisely and
to extract the most accurate information on the
parameter of the QCD symmetry from the observation.
Actually, in Ref.~\cite{Ikeno}, Ikeno {\it et al.} reported that 
we need very accurate
spectroscopic studies of the pionic atoms to deduce the density
dependence of the chiral condensate $\langle \bar{q} q \rangle$ 
from the observation beyond the linear density approximation~\cite{Goda}. 
It is also interesting and important to widen the domain of
the spectroscopic studies
of the deeply bound pionic states by including the odd-N target
for the systematic information on pion-nucleus systems and
the study of the reaction mechanism of the pionic atom formation
as in the case of analyzing
the reaction at finite angles~\cite{Ikeno2}.

In this article, we report the theoretical study of the ($d,^3$He) 
reaction for the pionic atom formation on the odd-N nuclear targets
$^{117}$Sn.



\section{Effective Number Approach}\label{Formula}
In the effective number approach in this article, 
we assume the neutron configuration of the ground state of the target
nucleus as, 
\begin{equation}
|i \rangle = |s_{1/2} \otimes 0^{+} \rangle,
\label{initial}
\end{equation}
by considering the odd-N nucleus 
with $J^{P}=\frac{1}{2}^{+}$ such as
$^{117}$Sn and $^{119}$Sn.
In Eq.~(\ref{initial}), we consider an extra neutron 
exists in the $s_{1/2}$ orbit together
with the $J^{P}=0^{+}$ even-even core nucleus in the target.
As for the final state, we distinguish two cases in our formula
which are written as,
\begin{eqnarray}
|f \rangle = \left\{
\begin{array}{l}
|\ell_{\pi} \otimes 0^{+} \rangle \\
|(\ell_{\pi} \otimes [s_{1/2} \otimes j_{h}^{-1}]_{J} )_{J_f} \rangle
\end{array}
\right.,
\label{final}
\end{eqnarray}
where we write the quantum number of the extra neutron state as $s_{1/2}$ and 
the neutron hole state as $j_h^{-1}$ with
the total angular momentum $J$ of the daughter nucleus.
The angular momentum of the pion bound state is written
as $\ell_\pi$ 
and the total angular momentum of the pionic atom as $J_f$.
In the first line in Eq.~(\ref{final}), 
we consider the final state with the
atomic pion and the even-even core nucleus in the ground state after the
extra $s_{1/2}$ neutron is removed from the target in the ($d,^3$He) reaction.
In the second line in Eq.~(\ref{final}), 
one neutron is considered to be picked-up from other
neutron state leaving a neutron-hole
$j_h^{-1}$ in the core of the daughter nucleus.

The contributions of these final states to the ($d,^3$He) spectra
are evaluated as the effective numbers 
$N_{\rm eff}^{(1)}$ and $N_{\rm eff}^{(2)}$ defined by,
\begin{eqnarray}
N_{\rm eff}^{(1)}
= \frac{1}{2} \sum_{m_n}    
\sum_{m_\pi} \sum_{m_s} 
\left| 
\int d \boldsymbol{r} e^{i \boldsymbol {q} \cdot \boldsymbol{r}} D(\boldsymbol{r})
\xi^{\dagger}_{\frac{1}{2} m_{s}} 
\langle \ell_{\pi} \otimes 0^{+}|
\hat{\phi}_{\pi}^{\dagger}(\boldsymbol{r}) \hat{\psi_{n}}(\boldsymbol{r})
|s_{1/2} \otimes 0^{+} \rangle
\right|^2,
\label{Neff_1}
\end{eqnarray}
and
\begin{eqnarray}
N_{\rm eff}^{(2)}
=
\frac{1}{2} \sum_{m_n} 
\sum_{J_f M_f} \sum_{m_s} 
\left| \int d \boldsymbol{r} e^{i \boldsymbol{q} \cdot \boldsymbol{r}} D(\boldsymbol{r})
\xi^{\dagger}_{\frac{1}{2} m_{s}} 
\langle (\ell_{\pi} \otimes [s_{1/2} \otimes j_{h}^{-1}]_{J} )_{J_f}|
\hat{\phi}^{\dagger}_{\pi}(\boldsymbol{r}) \hat{\psi_{n}}(\boldsymbol{r})
|s_{1/2} \otimes 0^{+} \rangle
\right|^2. \nonumber\\
\label{Neff_2}
\end{eqnarray}
$N_{\rm eff}^{(1)}$ is the effective number for the pionic atom
formation in the case of neutron pick-up
from the $s_{1/2}$ orbit
in the target nucleus and 
$N_{\rm eff}^{(2)}$ is that of neutron pick-up from 
a single particle orbit $j_h$ other
than $s_{1/2}$, which corresponds
to the final states shown in Eq.~(\ref{final}).
$\hat{\phi}_{\pi}^{\dagger}$ and
$\hat{\psi_{n}}$ indicate the field operators of pion and neutron
in analogous with the formula for the even-even nuclear 
target~\cite{nd_d3He,d3He}.
The spin wave function 
of the picked-up neutron
is denoted as $\xi_{\frac{1}{2} m_{s}}$, and we take
the spin average with respect to
$m_{s}$ so as to take into account the possible spin direction of the
neutrons in the target nucleus. 
We take the spin average of the neutron in the $s_{1/2}$ state 
in the target as indicated by
$\displaystyle \frac{1}{2} \sum_{m_n}$
in Eqs.~(\ref{Neff_1}) and~(\ref{Neff_2}).
We use the Eikonal approximation and 
write the projectile ($d$) distorted wave $\chi_d$ 
and the ejectile ($^3$He) distorted wave $\chi^{*}_{\rm He}$ as,
\begin{equation}
\chi^{*}_{\rm He}(\boldsymbol{r})\chi_{d}(\boldsymbol{r})=
e^{i \boldsymbol{q} \cdot \boldsymbol{r}} D(\boldsymbol{r}),
\label{D(z)}
\end{equation}
where 
$\boldsymbol{q}$ is the momentum transfer between the projectile 
and the ejectile,
and the distortion factor $D(\boldsymbol{r})$  is defined as
\begin{eqnarray}
D(\boldsymbol{r})=D(z, \boldsymbol{b})=
\exp
\Big[-\frac{1}{2}\sigma_{dN}\int_{-\infty}^{z}dz^{\prime}\rho_{A}(z^{\prime},\boldsymbol{b}) -
 \frac{1}{2}\sigma_{hN}\int_{z}^{\infty}dz^{\prime}\rho_{A-1}(z^{\prime},\boldsymbol{b})\Big]. 
\label{D(z)2}
\end{eqnarray}
Here, the deuteron-nucleon and $^3$He-nucleon total cross sections are
denoted as $\sigma_{dN}$ and $\sigma_{hN}$. 
The function $\rho_A(z,\boldsymbol{b})$ and
$\rho_{A-1}(z,\boldsymbol{b})$ are the density distributions of the
target and daughter nuclei at beam-direction coordinate $z$ with impact
parameter $\boldsymbol{b}$. 

The effective numbers $N_{\rm eff}^{(1)}$ and $N_{\rm eff}^{(2)}$
in Eqs.~(\ref{Neff_1}) and (\ref{Neff_2})
are manipulated separately and 
they are reduced to
\begin{eqnarray}
N_{\rm eff}^{(1)}
&=&
\sum_{m_\pi} 
\left|\int d \boldsymbol{r} e^{i \boldsymbol{q} \cdot \boldsymbol{r}} D(\boldsymbol{r})
\phi^{*}_{\ell_\pi m_\pi}(\boldsymbol{r})
\phi_{0 0}(\boldsymbol{r}) \right|^2 ,
\label{Neff_1.2}
\end{eqnarray}
and
\begin{eqnarray}
N_{\rm eff}^{(2)}
&=&
\frac{1}{2} \sum_{J_f j} \sum_{\ell n} 
\Bigg| \langle 6j \rangle \langle 6j' \rangle
\sqrt{\frac{2 J_{f} + 1} {2 \ell + 1}}
\sum_{m_\pi m_h}(\ell_\pi m_\pi  \ell_h m_h |\ell n)
(-)^{\ell_{h} - m_{h}} 
\nonumber\\
&\times&
\int d \boldsymbol{r} e^{i \boldsymbol{q} \cdot \boldsymbol{r}} D(\boldsymbol{r})
\phi^{*}_{\ell_\pi m_\pi}(\boldsymbol{r}) \phi_{\ell_h -m_h}(\boldsymbol{r})
\Bigg|^2 , 
\label{Neff_2.2}
\end{eqnarray}
respectively.
Here, $\phi_{\ell_\pi m_\pi}$ indicates 
the spatial wave function of the pion in the daughter nucleus 
and $\phi_{\ell_h m_h}$ that of
the neutron bound state in the target nucleus.
The pion wave function $\phi_{\ell_\pi m_\pi}$ is calculated by solving the
Klein-Gordon equation with the realistic potential
including the finite Coulomb potential 
and the optical potential~\cite{Piatom}.
For the neutron, 
we use the calculated spatial wave function $\phi_{\ell_h m_h}$ 
by the neutron potential reported in Ref.~\cite{koura}.

In Eq.~(\ref{Neff_2.2}),
the $\langle 6j \rangle$ symbol
indicates the coefficient of the expansion as
\begin{eqnarray}
\langle 6j \rangle 
&=& 
\langle (\ell_\pi j_{h})j s_{1/2} ;J_f
| \ell_\pi (j_{h} s_{1/2})J ;J_f \rangle
\nonumber\\
&=&\sqrt{(2j+1)(2J+1)} W(\ell_{\pi} j_{h} J_{f} \frac{1}{2}; j J)\nonumber\\
&=&\sqrt{(2j+1)(2J+1)} (-)^{-(\ell_{\pi}+j_{h}+J_{f}+\frac{1}{2})}
\left\{
\begin{array}{ccc}
\ell_{\pi}  & j_{h} & j  \\
\frac{1}{2} & J_{f}  & J
\end{array}
\right\},
\end{eqnarray}
where $W(\ell_{\pi} j_{h} J_{f} \frac{1}{2}; j J)$ is the Racah coefficient 
and 
$\left\{ \begin{array}{ccc}
\ell_{\pi}  & j_{h} & j  \\
\frac{1}{2} & J_{f}  & J
\end{array}
\right\}$
the 6$j$-symbol.
The symbol
$\langle 6j' \rangle$ 
also indicates the coefficient of the expansion
$\displaystyle \langle(\ell_{\pi} \ell_h)\ell\frac{1}{2}; j |
\ell_{\pi} (\ell_h \frac{1}{2}) j_{h} ;j\rangle $.

Using these effective numbers, the bound state formation cross section can
be written as,
\begin{eqnarray}  
\left( \frac{d^{2}\sigma}{dE_{\rm He} d\Omega_{\rm He}} 
\right)^{\rm lab}_{A} 
=
\left(\frac{d\sigma}{d\Omega_{\rm He}}\right )^{\rm lab}_{\rm ele}
\sum_{i=1}^{2}  
\sum_{\rm spins}  
K
\frac{\Gamma}{2\pi}\frac{1}{\Delta E^{2}+{\Gamma}^{2}/4}
N_{\rm eff}^{(i)}, 
\label{Cross_bound}
\end{eqnarray}
where 
the symbol of the sum of spin quantum numbers
$\displaystyle \sum_{\rm spins}$ indicates 
$\displaystyle \sum_{\ell_\pi}$ for $i=1$ and
$\displaystyle \sum_{\ell_\pi j_h J}$ for $i=2$, respectively.

Here,
$\displaystyle \left(\frac{d\sigma}{d\Omega_{\rm He}}\right )^{\rm lab}_{\rm ele}$
indicates the elementary differential cross section 
for the $d + n \rightarrow {^3{\rm He}} + \pi^-$ 
reaction in the laboratory system, which is extracted
from the experimental data~\cite{ele_data} of the 
$p + d \rightarrow t + \pi^+$ reaction assuming charge symmetry~\cite{Ikeno2}.
We use 
the Lorentz distribution function 
$\displaystyle \frac{\Gamma}{2\pi}\frac{1}{\Delta E^{2}+{\Gamma}^{2}/4}$
to account for the width
$\Gamma$ of the pion bound state, where 
$\Delta E$ is defined as $\Delta E=E_{\rm He} +E_\pi - E_d -E_n$. 
$E_n$ is the energy of neutron in the target nucleus and defined as
$E_n=M_n-S_n (j_n)$ with the neutron separation energy $S_n$ from the
$j_n$ single particle level and the neutron mass $M_n$.
$E_\pi$ is the energy of the pion bound state defined as 
$E_\pi=m_\pi-B.E.(\ell_\pi)$
with the pion mass $m_\pi$ and the binding energy $B.E.$ of the bound
state indicated by $\ell_\pi$.
The reaction $Q$-value can be expressed as
$Q=\Delta E - m_\pi + B.E.(\ell_\pi)-S_n (j_n) +(M_n + M_d -M_{\rm He})$, 
where $M_n + M_d -M_{\rm He}=6.787$ MeV.

The kinematical correction factor $K$ in Eq.~(\ref{Cross_bound}) is defined as
\begin{eqnarray}
K=\left[ \frac{|\boldsymbol{p}^A_{\rm He}|}{|\boldsymbol{p}_{\rm He}|} 
\frac{E_n E_\pi} {E^A_n E^A_\pi} 
\left(1+\frac{E_{\rm He}}{E_\pi}\frac{|\boldsymbol{p}_{\rm He}|-|\boldsymbol{ p}_d| 
{\rm cos}\theta_{d{\rm He}}}{|\boldsymbol{p}_{\rm He}|} \right) \right]^{\rm lab},
\label{K}
\end{eqnarray}
where the superscript `$A$' indicates the momentum and energy 
which are evaluated
in the kinematics of the nuclear target case~\cite{Ikeno2}.
The superscript `lab' indicates that all kinematical variables are
evaluated in the laboratory frame.
This correction factor is $K=1$ for the recoilless kinematics at 
$\theta_{d{\rm He}}^{\rm lab}=0^\circ$ with $S_n =0$ and $B.E.=0$.

As for the quasi-free contributions, we can express the cross
section as,
\begin{eqnarray}
\left(\frac{d^2 \sigma}{dE_{\rm He} d\Omega_{\rm He}}\right)^{\rm lab}_{A}
= 
\left(\frac{d\sigma}{d\Omega_{\rm He}}\right )^{\rm lab}_{\rm ele}
\sum_{i=1}^{2}  
\sum_{\rm spins}
\frac{2 |\boldsymbol{p}^{A}_\pi| E^{A}_\pi}{\pi} K N_{\rm eff}^{(i)}.
\label{Cross_qe}
\end{eqnarray} 
The definition of the symbol $\displaystyle \sum_{\rm spins}$ is the
same as in Eq.~(\ref{Cross_bound}).
The factor $\displaystyle \frac{2 |\boldsymbol{p}^{A}_\pi| E^{A}_\pi}{\pi} $
is due to the phase volume of the unbound pion~\cite{QE}.

In order to predict the realistic spectrum shape of the ($d$,$^3$He)
reactions for the pionic state formation, 
we need to take into account the realistic ground-state configurations
of the target nuclei, 
the nuclear excitation energies, and the relative excitation
strengths leading to the excited states of the daughter nuclei. 
First, we need to normalize the calculated effective numbers 
using the neutron occupation probability of each single particle state
in the ground state of the target nucleus
to obtain a realistic total strength for the neutron pick-up process
from each orbital.
The occupation probabilities are obtained from the analyses of the
data of the one neutron pick-up reactions such as ($p, d$) and
($d, t$) 
and are not equal to one in general.

As for the excited levels of the daughter nuclei, 
we use the experimental excitation energies and strengths 
obtained from the one neutron pick-up reactions.
%
For odd-N nuclear target case with $J^P =\frac{1}{2}^{+}$ such as $^{117}$Sn,
we consider two kinds of the final state separately as shown
in Eq.~(\ref{final}).
If a single neutron is picked-up from 
the neutron orbit $s_{1/2}$ in the target, 
the daughter nucleus is expected to have total angular momentum $J^P =0^+$.
In our formula, 
the effective number $N_{\rm eff}^{(1)}$
for the pionic state ($\ell_\pi$) formation with
such a state is modified as,
\begin{eqnarray}
N_{\rm eff}^{(1)}
\rightarrow 
N_{\rm eff}^{(1)}(\ell_{\pi}\otimes (J^P =0^+))
\times F_{O}(s_{1/2}) \times
\left\{
\begin{array}{l}
F_{R}((J^P =0^+)_1), \\
F_{R}((J^P =0^+)_2), \\
...\\
F_{R}((J^P =0^+)_N), \\
\end{array}
\right.
\label{Neff_FoFr_1}
\end{eqnarray}
where `$N$' indicates the number of states
of the daughter nucleus with $J^P =0^{+}$.
And $N_{\rm eff}^{(1)}(\ell_{\pi}\otimes (J^P=0^+))$
is the effective number defined in 
Eq. (\ref{Neff_1}),
$F_{O}$ the normalization factor
due to the occupation probabilities of the neutron states $s_{1/2}$ in the
target nucleus,
and $F_R$ the relative strength factor of the $N$-th excited state
in the daughter nucleus with total angular momentum $J^P =0^+$~\cite{Umemoto,Umemoto2}.
The contributions from the all $J^P =0^+$ nuclear excited states are 
obtained by summing up the r.h.s of Eq.~(\ref{Cross_bound}) 
with the realistic (experimental) excitation energies
which appeared in $\Delta E$.

Then, if a single neutron is picked-up from other neutron orbit $j_h$,
the daughter nucleus has the total angular momentum 
$J = [s_{1/2} \otimes j_h^{-1} ]$.
The effective number $N_{\rm eff}^{(2)}$
for the pionic state ($\ell_\pi$) formation with
the excited state of the daughter nucleus with 
$J= [s_{1/2} \otimes j_h^{-1} ]$
is modified in our model as,
\begin{eqnarray}
N_{\rm eff}^{(2)}
\rightarrow 
N_{\rm eff}^{(2)}(\ell_{\pi}\otimes [s_{1/2} \otimes j_h^{-1} ]_J)
\times F_{O}(j_{h}) \times 
\left\{
\begin{array}{l}
F_{R}(([s_{1/2} \otimes j_h^{-1} ]_J)_1),\\
F_{R}(([s_{1/2} \otimes j_h^{-1} ]_J)_2),\\
...\\
F_{R}(([s_{1/2} \otimes j_h^{-1} ]_J)_N),\\
\end{array}
\right.
\label{Neff_FoFr_2}
\end{eqnarray}
where `$N$' indicates the number of states
of the daughter nucleus with spin $J$.
And $N_{\rm eff}^{(2)}(\ell_{\pi}\otimes [s_{1/2} \otimes j_h^{-1} ]_J)$
is the effective number defined in 
Eq. (\ref{Neff_2}),
$F_{O}$ the normalization factor
due to the occupation probabilities of the neutron states $j_h$ in the
target nucleus, and
$F_R$ the relative strength factor of the $N$-th excited states
in the daughter nucleus with total angular momentum 
$J$ with a $j_h^{-1}$ neutron hole~\cite{Umemoto,Umemoto2}.
The contributions from all the nuclear excited states with total angular
momentum $J$ are obtained by summing up them in Eq.~(\ref{Cross_bound}) 
as in the case with $J^P =0^+$ described above.

We show in the next section the concrete examples of the application of
the $F_O$ and $F_R$ 
obtained from the $^{117}$Sn$(d, t) ^{116}$Sn reaction~\cite{Exp_FoFr}
to calculate the spectra of the $^{117}$Sn($d,^3$He) reactions for the
pionic atom formation.

\section{Numerical Result}
We show the calculated results of the $^{117}$Sn($d,^3$He) reaction
spectra for the formation of the deeply bound pionic states in
$^{116}$Sn, and compare the spectra with that of the $^{122}$Sn nuclear
target case.
We determine firstly 
the normalization factors $F_O$ of the neutron states in the target nucleus
and the relative strengths $F_R$ of the nuclear excited levels of the
daughter nucleus 
using the spectroscopic strength obtained from 
$^{117}$Sn($d, t$)$^{116}$Sn data~\cite{Exp_FoFr}.

In the Table 2 of Ref.~\cite{Exp_FoFr},
we can find the compilation of the data of the excitation energy $E_x$,
the spin-parity $J^{P}$ of the final nuclear state,
and the spectroscopic strength $G(j_h)$ of the the neutron hole state $j_h$.
As we can see from the table, the quantum numbers of the some of the
excited states could not be determined by the analyses of 
the experiment in Ref.~\cite{Exp_FoFr}.
The spin-parity $J^P$ is missing for the levels of $E_x=$
3.228, 3.315, 3.371, 3.470, 3.513, 3.589, 3.618, 3.772, 3.950, 
4.037, 4.084 MeV,
and 
the spectroscopic strength $G(j_h)$ of the neutron hole state
is not uniquely determined for $J^P =2^+$ states
which are written as $G$($d_{3/2}$ or $d_{5/2}$) in the table.
The spin-parity is not uniquely determined for the level of 
$E_x=$ 3.739 MeV.

In order to make use of these data~\cite{Exp_FoFr}
to determine the $F_O$ and $F_R$ factors,
we need to make a few assumptions
to compensate for the missing information 
in the Table 2 of Ref.~\cite{Exp_FoFr}.
First, we assume 
the value of the missing and not-uniquely determined
($E_x=$ 3.739 MeV)
spin-parity of the states 
listed in the column of $G$($d_{3/2}$ or $d_{5/2}$)  
to be $J^P =2^+$
since it can be realized for both $d_{3/2}$ and $d_{5/2}$ states by
coupling to the $s_{1/2}$ state.
Then, we assume that the spectroscopic strength of the levels,
for which the observed strengths are
written in the {\it both} columns of $d_{3/2}$ and $d_{5/2}$ of 
$G$($d_{3/2}$ or $d_{5/2}$), 
is equally shared by the contributions for $j_h=d_{3/2}$ and $d_{5/2}$ states.
For example, 
the spectroscopic strength $G$($d_{3/2}$ or $d_{5/2}$) 
at $E_x =$1.294 MeV is written as 
$G(d_{3/2})=$ 0.20 and
$G(d_{5/2})=$ 0.16 in Table 2 in Ref.~\cite{Exp_FoFr}.
This means that the observed strength corresponds to $G=0.20$ in
case of $j_h =d_{3/2}$, and $G=0.16$ for $j_h=d_{5/2}$ case.
We assume that
there exist the half contributions from both $j_h$ as
$G(d_{3/2})=0.10$ and $G(d_{5/2})=0.08$.
We determine all spectroscopic strength 
$G(d_{3/2})$ and $G(d_{5/2})$ in a similar way.
%
We also assume that the spin-parity of the level at $E_x=$ 3.315 MeV 
is $J^P =3^+$.
Since the strength of this level is small, this assumption will not
affect the final results.
Finally, 
we neglect the level of $f_{5/2}$ or $f_{7/2}$
in $E_x =2.266$ MeV in Table 2 of Ref.~\cite{Exp_FoFr}
since the strength of these excitations is small and 
they correspond to the core
polarization effects in the naive shell model.

Based on the assumptions described above,
we use the data in Ref.~\cite{Exp_FoFr} to determine 
the $F_O$ and $F_R$ factors.
In Table~\ref{117Sn_Fo},
we summarize the normalization factors $F_O$
corresponding to the occupation probabilities of the neuron orbit $j_h$ in the
target $^{117}$Sn.
Since the sum of the spectroscopic strength $G(j_h)$ in Ref.~\cite{Exp_FoFr}
is expected to be equal to the number of the neutron in 
the neutron orbit ($j_h$) in the target,
we evaluate $F_O(j_h)$ as, 
\begin{equation}
F_O(j_h) =  \frac{1}{(2 j_h + 1)} \displaystyle 
\sum_{N} G(j_h)_N,
\label{Fo1}
\end{equation}
where $ G(j_h)_N$ indicates the spectroscopic strength of the $N$-th nuclear
levels with a hole state $j_h$ in the daughter nucleus.
As for $j_h = s_{1/2}$ case, 
we evaluate $F_O(s_{1/2})$ as, 
\begin{equation}
F_O(s_{1/2}) =  \displaystyle \sum_{N} G(s_{1/2})_N,
\label{Fo2}
\end{equation}
since we assumed that 
the number of the neutron in the $s_{1/2}$ orbit in the odd-N target nucleus
with $J^P = \frac{1}{2}^+$
is equal to be 1 for fully occupied case
as shown in Eq.~(\ref{initial})
in the effective number formalism in Section~\ref{Formula}.

\begin{table}[!tb]
\caption{
Compilation of 
the normalization factor ($F_O$) corresponding to the occupation 
probability of each neutron state in the ground state of $^{117}$Sn
determined from the experimental data in Ref.~\cite{Exp_FoFr}.
See details in the text.}
\label{117Sn_Fo}
\centering
\begin{tabular}{cccc}
\hline
Neutron hole orbit ($j_h$) & Normalization factor ($F_{O}$) \\ 
\hline
$3s_{1/2}$   & 0.47 \\
$2d_{3/2}$   & 0.40 \\
$2d_{5/2}$   & 0.39 \\
$1g_{7/2}$   & 0.80 \\
$1h_{11/2}$  & 0.11 \\ 
\hline
\end{tabular}
\end{table}

In Table~\ref{117Sn_Fr},
we summarize the spin-parity $J^{P}$,
the excitation energies $E_x$,
the neutron hole orbit $j_h$,
and the relative strength $F_R$ of the excited states of 
$^{116}$Sn determined from the data of $^{117}$Sn($d, t$)$^{116}$Sn
in Ref.~\cite{Exp_FoFr} with some assumptions described before.
We determine the relative strength $F_R$
using the spectroscopic strength $G(j_h)$ 
in Ref.~\cite{Exp_FoFr} again.
We evaluate $F_R(([s_{1/2} \otimes j_h^{-1} ]_J)_N)$ as, 
\begin{equation}
F_R(([s_{1/2} \otimes j_h^{-1} ]_J)_N) =  
\frac{G(j_h)_{N, J}}
{\displaystyle \sum_{N} G(j_h)_{N, J}},
\label{Fr1}
\end{equation}
where $G(j_h)_{N, J}$ indicates the spectroscopic strength of the $N$-th nuclear
levels with a hole state $j_h$ and total spin $J$ in the daughter nucleus.
For $j_h = s_{1/2}$ case,
we evaluate $F_R((J^P =0^+)_N)$ as, 
\begin{equation}
F_R((J^P=0^+)_N) =  
\frac{G(s_{1/2})_{N, 0^+}}
{\displaystyle \sum_{N} G(s_{1/2})_{N, 0^+}}.
\label{Fr2}
\end{equation}

\begin{table}[!tb]
\caption{
Excitation energy ($E_{x}$) and relative strength ($F_{R}$) of
 each excited level in $^{116}$Sn determined from the experimental data of
the $^{117}$Sn($d,t$) reactions~\cite{Exp_FoFr}. 
See the text for details.
}
\label{117Sn_Fr}
\centering
\begin{tabular}{cccccc}
\hline
 $J^{P}$&  $E_x$ [MeV] & Neutron hole orbit $j_h$ & $F_{R}$ &
 Neutron hole orbit $j_h$ &$F_{R}$ \\
\hline
$0^+$	& 0.000	&$3s_{1/2}$	& 0.68	&		&	\\
	& 1.757	&		& 0.13	&		&	\\
	& 2.027	&		& 0.11	&		&	\\
	& 2.545	&		& 0.09	&		&	\\
\hline
$1^+$	& 2.587	&$2d_{3/2}$	& 0.92	&		&	\\
	& 2.960	&		& 0.08	&		&	\\
\hline											
$2^+$	& 1.294	&$2d_{3/2}$	& 0.07	&$2d_{5/2}$ 	& 0.08	\\
	& 2.112	&		& 0.00	&		& 0.00	\\
	& 2.225	&		& 0.09  &		& 0.09	\\
	& 2.650	&		& 0.00	&		& 0.00	\\
	& 2.843	&		& 0.03	&		& 0.03	\\
	& 3.228	&		& 0.09	&		& 0.09	\\
	& 3.371	&		& 0.09	&		& 0.09	\\
	& 3.470	&		& 0.10	&		& 0.10	\\
	& 3.513	&		& 0.03	&		& 0.03	\\
	& 3.589	&		& 0.14	&		& 0.14	\\
	& 3.618	&		& 0.01	&		& 0.01	\\
	& 3.739	&		& 0.03	&		& 0.03	\\
	& 3.772	&		& 0.14	&		& 0.14	\\
	& 3.950	&		& 0.13	&		& 0.13	\\
	& 4.037	&		& 0.03	&		& 0.03	\\
	& 4.084	&		& 0.02	&		& 0.02	\\
\hline											
$3^+$	& 2.997	&$2d_{5/2}$	& 0.05	&$1g_{7/2}$	& 0.26	\\
	& 3.180	&		&	&		& 0.43	\\
	& 3.315	&		&	&		& 0.02	\\
	& 3.416	&		& 0.37	&		& 0.06	\\
	& 3.709	&		& 0.59	&		& 0.23	\\
\hline											
$4^+$	& 2.390	&$1g_{7/2}$	& 0.08	&		&	\\
	& 2.529	&		& 0.09	&		&	\\
	& 2.801	&		& 0.21	&		&	\\
	& 3.046	&		& 0.15	&		&	\\
	& 3.096	&		& 0.48	&		&	\\
\hline											
$5^-$	& 2.366	&$1h_{11/2}$	& 1.00	&		&	\\
\hline	
$6^-$	& 2.773	&$1h_{11/2}$	& 1.00	&		&	\\
\hline											
\end{tabular}
\end{table}

\begin{table}[tb]
\caption{
Excitation energy ($E_{x}$) and relative strength ($F_{R}$) of
each excited level in $^{121}$In determined from the experimental 
data of the $^{122}$Sn($d,^3$He)$^{121}$In reaction~\cite{122Sn_pi0}.
The normalization factor ($F_O$) corresponding to the occupation 
probability of each proton state in the ground state of $^{122}$Sn
is also listed.
Here, we have assumed the level at $E_x =1.40$ MeV to be $1f_{5/2}$ state,
which was not identified clearly in Ref.~\cite{122Sn_pi0}.
}
\label{122Sn_pi0}
\centering
\begin{tabular}{cccc}
\hline
Proton hole orbit ($j_h$) & $E_{x}$ [MeV] & Relative strength ($F_{R}$) 
& Normalization factor ($F_{O}$) \\ 
\hline
$1g_{9/2}$  & 0.00  & 1.0  & 1.0  \\
$2p_{1/2}$  & 0.31  & 1.0  & 1.0  \\
$2p_{3/2}$  & 0.62  & 1.0  & 1.0  \\
$1f_{5/2}$  & 1.40  & 1.0  & 1.0  \\ 
\hline
\end{tabular}
\end{table}

\begin{figure}[!tb]
\centering
\includegraphics[width=10.0cm,height=13.5cm]{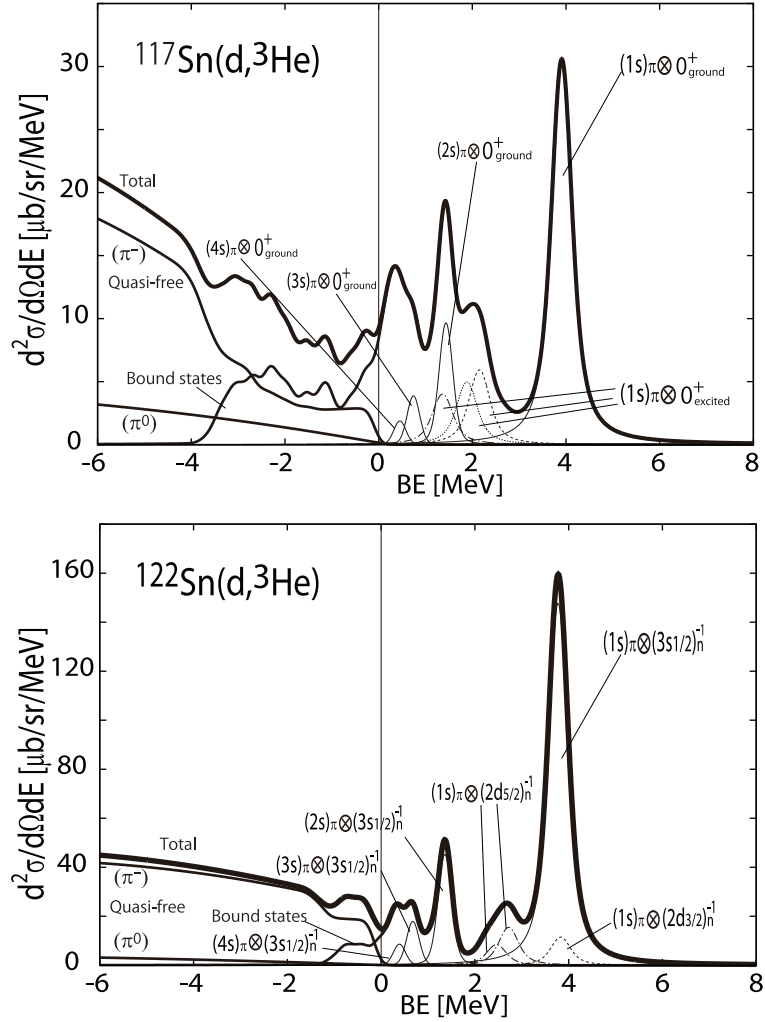}%
\caption{
Calculated spectra for the formation of the pionic states
in the $^{117}$Sn($d, ^{3}$He) (upper) and 
the $^{122}$Sn($d, ^{3}$He) (lower) reactions
at $\theta_{d{\rm He}}^{\rm lab}=0^\circ$ 
plotted as functions of the pion binding energy.
The incident deuteron kinetic energy is fixed to be $T_d = 500$ MeV.
The total spectra are shown by the thick solid lines in both figures.
We show separately
the contributions of the bound pionic states formation and 
the quasi-free $\pi^-$ and $\pi^0$ production by the thin solid lines.
The dominant subcomponents are also shown in the figures with 
quantum numbers indicated as
$[(n \ell)_{\pi} \otimes J^P]$
in the $^{117}$Sn($d,^3$He) reaction (upper)
and
$[(n \ell)_{\pi} \otimes (n \ell_j)^{-1}_n]$
in the $^{122}$Sn($d,^3$He) reaction (lower), respectively.
The instrumental energy resolution is assumed to be 300 keV FWHM.
The contribution of the $\pi^0$ quasi-free production in the 
$^{117}$Sn($d,^3$He) reaction (upper) is assumed to be the same as that
of $^{122}$Sn($d,^3$He) reaction (lower)
except for the shift of the threshold energy.
See the text for details.
}
\label{Cross_0deg}
\end{figure}

In Fig.~\ref{Cross_0deg}, we show the calculated spectra
for the pionic states formation
at $\theta_{d{\rm He}}^{\rm lab}=0^\circ$ 
in the $^{117}$Sn($d,^{3}$He) and $^{122}$Sn($d,^{3}$He) reactions.
In addition to the total spectra,
the contributions of the deeply bound pionic states formation,
and
the quasi-free $\pi^-$ and $\pi^0$ production are also shown separately.
The dominant subcomponents are indicated in the figure
with their quantum numbers.
In the $^{117}$Sn($d,^{3}$He) spectra,
the contributions from the final states $[(n \ell)_{\pi} \otimes J^P]$
with $J^P =0^+$ and $\ell_\pi=0$ are dominant.
We find that we can see clearly the peak structure 
of the pionic 1$s$ state formation 
with the ground state of the even-even nucleus $^{116}$Sn
as indicated in the figure as [$(1s)_{\pi} \otimes 0^{+}_{\rm ground}$].
The subcomponents coupled to the ground state of the daughter nucleus 
$^{116}$Sn in the upper figure of Fig.~\ref{Cross_0deg}
will not have the additional shifts due to the residual
interaction effects. 
In the $^{122}$Sn($d,^{3}$He) spectra (lower figure),
the cross section in the bound pion region is the same as in Ref.~\cite{Ikeno2}.

In both $^{117}$Sn($d,^{3}$He) and $^{122}$Sn($d,^{3}$He) spectra, 
the contribution from the pionic 1$s$ state formation with 
the neutron $s$ hole state is found to be large
because of the matching condition with the recoilless kinematics. 
We find that the total spectrum of the bound pionic state formation 
in $^{117}$Sn($d,^{3}$He) reaction
spreads into wider energy range
than that in $^{122}$Sn($d,^{3}$He) reaction.
This is because 
the excited levels of $^{116}$Sn 
tend to have larger excitation energies ($E_{x}$) 
than those of $^{121}$Sn~\cite{Ikeno}.

As shown in Fig.~\ref{Cross_0deg},
the $\pi^0$ quasi-free production contribution 
is found to be much smaller than that of $\pi^-$.
Here, the contribution of the quasi-free $\pi^0$ production 
in the $^{117}$Sn($d,^3$He) spectra is estimated
approximately as explained below.

To calculate the contributions of the quasi-free $\pi^0$ production,
we use the experimental data of the single proton pick-up reactions
from the same target to determine the excitation
energies and strengths of the one proton-hole states of the nucleus
as in the case of $\pi^-$ production with a neutron-hole.
In the $^{117}$Sn target case, 
we do not find
the appropriate experimental data of the single proton pick-up reactions.
Therefore,
as the quasi-free $\pi^0$ production contributions
of the $^{117}$Sn target case, 
we assume 
the same as those of the $^{122}$Sn target except the shift of
the threshold energy.
We consider that this assumption does not affect 
the essential structure of the total spectrum
since the contribution of the quasi-free $\pi^{0}$ production is 
much smaller than that of the quasi-free $\pi^-$ production 
around the threshold,
and is rather structureless except for gradual increment from the threshold.
It seems also reasonable
because of the expected similarities of the structure of the proton
states in the isotopes.
In Table~\ref{122Sn_pi0},
we summarize the excitation energies $E_x$,  the relative strengths $F_R$,
and the normalization factors $F_O(j_h)$ 
obtained from the experimental data
of the $^{122}$Sn($d,^3$He)$^{121}$In reaction~\cite{122Sn_pi0},
which are used to calculate the quasi-free $\pi^0$ production
contributions in the $^{122}$Sn($d, ^3$He) reaction.

\begin{figure}[!tb]
\centering
\includegraphics[width=15.0cm,height=5.5cm]{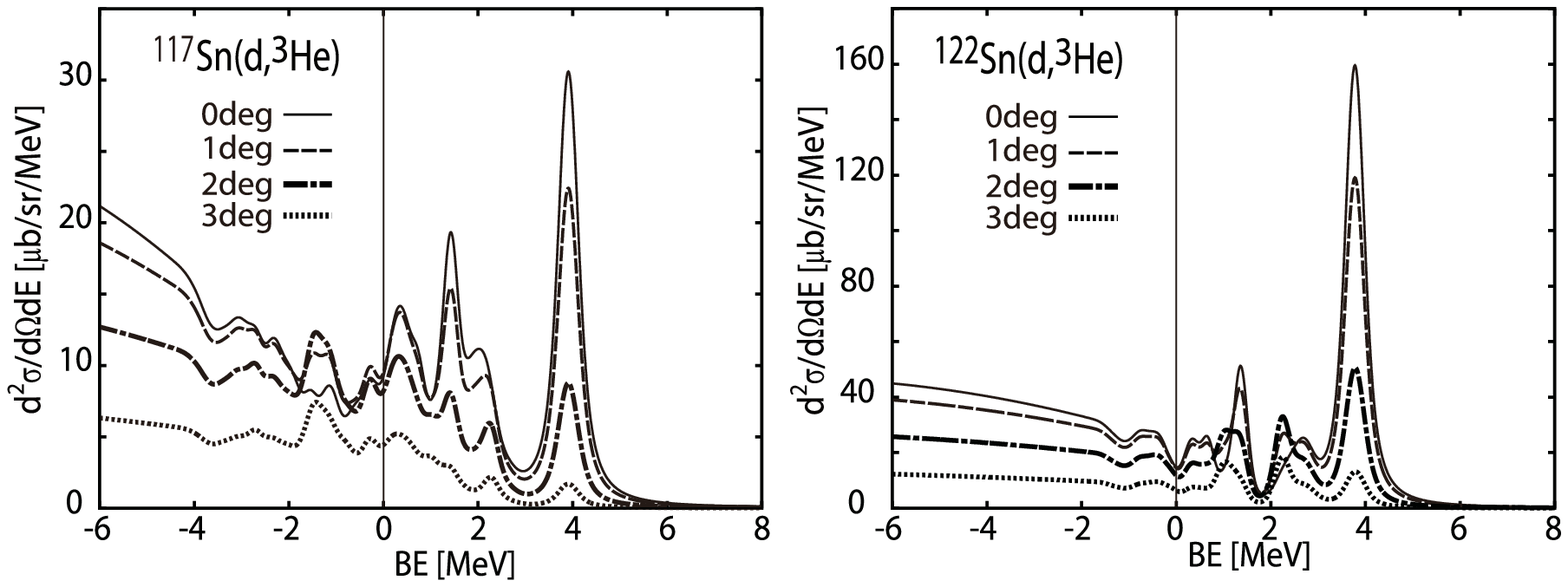}%
\caption{
Calculated spectra for the formation of the pionic states 
at $\theta^{\rm lab}_{d{\rm He}}=0^\circ$ (solid lines), 
$1^\circ$(dash lines), 
$2^\circ$ (dash-dotted lines) and $3^\circ$ (dotted lines) 
in the $^{117}$Sn($d, ^{3}$He) (left) and 
the $^{122}$Sn($d, ^{3}$He) (right) reactions
plotted as functions of 
the pion binding energy.
The incident deuteron kinetic energy is fixed to be $T_d = 500$ MeV.
The instrumental energy resolution is assumed to be 300 keV FWHM.
}
\label{Cross}
\end{figure}

In Fig.~\ref{Cross}, 
we show the calculated spectra for the formation 
of the pionic states at finite angles in the
$^{117}$Sn($d, ^{3}$He) and $^{122}$Sn($d, ^{3}$He) reactions.
We find that both spectra
have a strong angular dependence
because of the matching condition of the reaction
by which the different subcomponents dominate the spectra at different angles.
We also find that 
the absolute value of the calculated cross sections
in $^{117}$Sn($d, ^{3}$He) reaction
are significantly smaller than that in $^{122}$Sn($d,^{3}$He) reaction
for all cases considered in this article.
In Appendix~\ref{Appendix_A}, 
we make some remarks about the difference between the ($d,^3$He) spectra 
with $^{117}$Sn and $^{122}$Sn targets.

\section{Conclusion}
We study the ($d,^3$He) reaction for the pionic atom formation 
on the odd-N nuclear target theoretically.
In this reaction,
we can expect to observe the pionic states in the even-even nucleus 
with $J^P = 0^+$. 
These pionic states populated with the ground state of the daughter nucleus
could not be affected by
the additional shifts due to the residual interaction effects
between the neutron-hole and pionic states.
The experimental error of the latest data~\cite{KSuzuki} is significantly
smaller than 
that of the old ones~\cite{Itahashi,Geissel} 
and is now approaching the magnitude of
the calculated shifts due to the residual interaction effects~\cite{Kaneyasu,Nose}.
Thus, the observation of the pionic states free from these effects
is very important to obtain more accurate information on pion properties
in nucleus from the data, which
will be necessary to deduce the density dependence of the chiral condensate 
$\langle \bar{q} q \rangle$ 
beyond the linear density approximation~\cite{Ikeno,Goda}.

We extend the effective number approach to calculate
the pionic atom formation spectra on the odd-N nuclear target.
In this formulation,
we assume that
the target nucleus with $J^P = \frac{1}{2}^+$  
is composed of the even-even core nucleus with $J^{P}=0^{+}$
and the extra neutron located in the $s_{1/2}$ orbit.
Based on this assumption,
we consider that the pionic atom in the final state is populated in
the even-even core nucleus
in the case of single neutron pick-up from the $s_{1/2}$ orbit
in the target nucleus in the ($d,^3$He) reaction.
On the other hand,
we consider the pionic atoms 
are populated in the daughter nucleus with the configuration 
$J= [s_{1/2} \otimes j_h^{-1} ]$,
where the neutron remaining in the valence $s_{1/2}$ orbit is coupled 
to a neutron-hole state $j_h^{-1}$,
in the case of neutron pick-up from a single particle orbit $j_h$ other 
than $s_{1/2}$ in the target. 
We evaluate separately the effective numbers 
for these two kinds of final states
and obtain the pionic atom formation spectra by summing up all
contributions.

In order to predict the realistic spectrum shape of 
the ($d$,$^3$He) reaction, 
we take into account 
the normalization factor ($F_O$) of the neutron state
in the target nucleus
and the relative strength ($F_R$) of the nuclear excited levels
of the daughter nucleus as in Refs.~\cite{Umemoto,Umemoto2}, 
which are determined using the spectroscopic strength obtained 
from the experimental data~\cite{Exp_FoFr}.

We show the numerical results of $^{117}$Sn($d,^3$He) spectra 
for the pionic atom formation.
We find that we can see clearly
the peak structure of the pionic 1$s$ state formation 
with the ground state of the daughter nucleus $^{116}$Sn
which does not have the additional shifts due to the residual interaction
effects.
By comparing the $^{117}$Sn($d,^3$He) spectra with
those of the $^{122}$Sn($d,^3$He),
we find that 
the bound pionic state formation spectra
in $^{117}$Sn($d,^{3}$He) reaction spread over wider energy range
than those in $^{122}$Sn($d,^{3}$He) reaction 
because of the larger excitation energies of $^{116}$Sn than 
those of $^{121}$Sn.
We also find that the absolute values of the calculated cross sections
in $^{117}$Sn($d,^{3}$He) reaction
are significantly smaller than those in $^{122}$Sn($d,^{3}$He) reaction
because the values of 
the normalization factor $F_O$ and the relative strength $F_R$
of $^{117}$Sn are smaller than those of $^{122}$Sn
as described in detail in Appendix~\ref{Appendix_A}.

As for the contributions of the subcomponents,
we find that
the [$(1s)_\pi \otimes 0^{+}_{\rm ground}$] subcomponent
dominates the largest peak structure appeared in the spectra
around $B.E. \simeq 3.9$ MeV in Fig.~\ref{Cross_0deg}
in the $^{117}$Sn target case,
whereas   
two subcomponents of
[$(1s)_\pi \otimes (3s_{1/2})^{-1}_n$] and 
[$(1s)_\pi \otimes (2d_{3/2})^{-1}_n$]
are included in $^{122}$Sn target case
because of the small difference ($\sim$60keV) 
between the separation energies of these two neutron levels
in $^{121}$Sn~\cite{Ikeno}.
In general, the peak in the spectra dominated by a single subcomponent is
known to be the most useful to determine the pion binding energy
and width 
because of the absence of the ambiguities in
the relative strength of the subcomponents.

In summary,
the odd-N nuclear target with $J^P =\frac{1}{2}^{+}$ 
is expected to be suited for the pionic 1$s$ state observation 
because 
the calculated spectra for $^{117}$Sn target clearly
show the isolated peak structure of 
the pionic 1$s$ state formation with the ground state 
of the even-even nucleus ([$(1s)_\pi \otimes 0^{+}_{\rm ground}$]),
having no residual interaction effects. 

Thus,
the formation of the pionic 1$s$ state by the ($d,^3$He) reactions 
on the odd-N nuclear target 
is preferable to extract the most accurate information on the
parameter of the QCD symmetry from the observation.
However,
we need to be careful for
the absolute values of the cross sections for the $^{117}$Sn target
case,
which are expected to be
smaller than those of $^{122}$Sn target.

The experiment for the pionic atom formation on the odd-N nuclear target
will be performed at RIBF/RIKEN in near future~\cite{RIBF}.
We think that 
these studies will provide the systematic information on
the pionic bound states in various nuclei 
and will help to develop the study of the pion properties and the
partial restoration of chiral symmetry in nuclei.
We believe that our results provide a good motivation for further
experimental studies.

\section*{Acknowledgment}
We acknowledge
the fruitful discussions from the experimental side
with K. Itahashi, T. Nishi and H. Fujioka.
We thank H. Nakada 
for the useful comments 
on the structure of the ground state of the odd-neutron nucleus.
We also thank D. Jido for the useful discussions. 
N. I. appreciates the support by the Grant-in-Aid for JSPS Fellows (No. 23-2274).
This work is partly supported by the Grants-in-Aid for Scientific Research 
(No. 24105707 and No. 24540274).


%

\vfill\pagebreak

\appendix
\section{Comparison of the ($d,^3$He) spectra of the pionic atom formation on the $^{117}$Sn and $^{122}$Sn targets}\label{Appendix_A}
In this appendix, we discuss 
the origin of the differences of the calculated cross sections
of the pionic atom formation in the ($d,^{3}$He) reaction on 
the $^{117}$Sn and $^{122}$Sn target cases.
As shown in Fig.~\ref{Cross_0deg}, 
the cross section of $^{117}$Sn($d,^{3}$He)
is about one over five of that of $^{122}$Sn($d,^{3}$He)
at the largest peak position at $B.E. \simeq 3.9$ MeV.
We think that
this difference is large
even if we take into account the fact that
the spectra of one-neutron pick-up process from $^{117}$Sn
spread over wider energy region than
those from $^{122}$Sn 
due to the larger excitation energies of the daughter nucleus
as shown in Table~\ref{117Sn_Fr}.

\begin{table}[!tb]
\caption{
Symbols used in this Appendix and definitions
of the effective number ($N_{\rm eff}$), the normalization factor ($F_{O}$), 
and the relative strength ($F_R$)
are summarized for
the odd- and even-neutron nuclear target cases. 
As for the odd-N nuclear target with $J^P =\frac{1}{2}^+$,
we distinguish two cases of the final state of the ($d,^3$He) reaction
and treat them separately.
See details in the text.
}
\label{Odd_Even}
\centering
\begin{tabular}{|c||c|c||c|}
\hline 
Target nucleus & \multicolumn{2}{c||}{Odd nucleus with $J^P =\frac{1}{2}^{+}$}
& Even nucleus \\
 & \multicolumn{2}{c||}{ (ex. $^{117}$Sn)} & (ex. $^{122}$Sn) \\
\hline \hline
Neutron state & $s_{1/2}$ & $j_h$ & $j_h$  \\
in target& & &  \\
\hline
Spin of daughter & $0$ & $[s_{1/2} \otimes j_h^{-1} ]_J$ & $j_h^{-1}$  \\
 nucleus & & &  \\
\hline
Effective & $N_{\rm eff}^{\rm odd (1)}(\ell_{\pi}\otimes (J^P =0^+))$ &
$N_{\rm eff}^{{\rm odd}{(2)}}(\ell_{\pi}\otimes [s_{1/2} \otimes j_h^{-1}]_J)$
&$N_{\rm eff}^{\rm even}(\ell_{\pi}\otimes j_{h}^{-1})$ \\
number ($N_{\rm eff}$) & (Eq.~(\ref{Neff_1})) & (Eq.~(\ref{Neff_2})) &  \\
\hline
Normalization & $F_O^{\rm odd}(s_{1/2})$ & $F_O^{\rm odd}(j_h)$ &  $F_O^{\rm even}(j_h)$ \\
 factor ($F_{O}$) & (Eq.~(\ref{Fo2})) & (Eq.~(\ref{Fo1})) & 		\\
\hline						
Relative & 
$F_R^{\rm odd}((J^P =0^+)_N)$ & $F^{\rm odd}_R(([s_{1/2} \otimes j_h^{-1} ]_J)_N)$ &
$F_R^{\rm even}( (j_h^{-1} )_N)$ \\
strength ($F_R$) & (Eq.~(\ref{Fr2})) & (Eq.~(\ref{Fr1})) & 		\\
\hline											
\end{tabular}
\end{table}

In the effective number approach used to evaluate
the pionic atom formation cross section 
in this article,
the size of the formation cross section of the pionic states
is described by the effective number $N_{\rm eff}$ 
defined in Eqs.~(\ref{Neff_1}) and (\ref{Neff_2}),
and the additional factors $F_O$ and $F_R$ 
introduced in
Eqs.~(\ref{Neff_FoFr_1}) and~(\ref{Neff_FoFr_2}). 
Thus, we discuss the difference of these parts separately.

In Table~\ref{Odd_Even},
we first summarize the symbols used in this Appendix and definitions of
the effective number ($N_{\rm eff}$), the normalization factor ($F_{O}$)
and the relative strength ($F_R$)
for the odd-neutron (odd-N) and even-neutron (even-N) nuclear target cases.
As for the odd-N nuclear target with $J^P =\frac{1}{2}^+$,
we distinguish two cases of the final state of the ($d,^3$He) reaction
and treat them separately
as shown in Eq.~(\ref{final}) in Section~\ref{Formula}. 
In the case of neutron pick-up
from the $s_{1/2}$ orbit
in the target nucleus,
the pionic atoms are populated in
the even-even core nucleus with $J^P =0^+$.
We express in this case the effective number as
$N_{\rm eff}^{\rm odd (1)}(\ell_{\pi}\otimes (J^P =0^+))$,
the normalization factor as $F_O^{\rm odd}(s_{1/2})$, and
the relative strength as $F_R^{\rm odd}((J^P =0^+)_N)$,
which are defined
by Eqs.~(\ref{Neff_1}),~(\ref{Fo2}) and~(\ref{Fr2}),
respectively.
In the case of neutron pick-up from a single particle orbit $j_h$ other than
$s_{1/2}$ in the target,
the pionic atoms are populated in
the nucleus with the configuration $J= [s_{1/2} \otimes j_h^{-1}]$,
where the neutron remaining in the valence $s_{1/2}$ orbit is
coupled to the neutron-hole state $j_h^{-1}$
to form the total nuclear spin $J$.
We express in this case the effective number as
$N_{\rm eff}^{{\rm odd}{(2)}}(\ell_{\pi}\otimes [s_{1/2} \otimes j_h^{-1} ]_J)$,
the normalization factor as $F_O^{\rm odd}(j_h)$, and
the relative strength as $F^{\rm odd}_R(([s_{1/2} \otimes j_h^{-1} ]_J)_N$,
which are defined
by Eqs.~(\ref{Neff_2}),~(\ref{Fo1}) and~(\ref{Fr1}), respectively.
As for the even-N nuclear target case,
the pionic atoms are populated in the nucleus with
a neutron-hole state $j_{h}^{-1}$.
The effective number $N_{\rm eff}^{\rm even}(\ell_{\pi}\otimes j_{h}^{-1})$, 
the normalization factor as $F_O^{\rm even}(j_h)$, and
the relative strength as $F_R^{\rm even}( (j_h^{-1} )_N)$ 
for the even-N nuclear target case
are defined in Refs.~\cite{Umemoto,Umemoto2,Ikeno}.

In the considerations below,
we postulate that the differences of the wave functions
and distortion factors between $^{117}$Sn and $^{122}$Sn targets
are small and can be neglected safely.
In this case, it is known that 
the relative size of the effective numbers are determined by
the multiplicity of the states and satisfy the relation,
for example,
\begin{equation}
N_{\rm eff}^{\rm odd (2)}(\ell_{\pi}\otimes [s_{1/2} \otimes j_h^{-1} ]_{J_1})
:
N_{\rm eff}^{\rm odd (2)}(\ell_{\pi}\otimes [s_{1/2} \otimes j_h^{-1}]_{J_2})
=(2J_{1}+1):(2J_{2}+1).
\end{equation}

We consider first the pionic atom formation
in the case with a neutron picked-up from the $s_{1/2}$ orbit
in the target nucleus.
For the odd-N nuclear $^{117}$Sn target case,
the effective number for 
the pionic atom ($\ell_\pi$) formation 
in the even-even core nucleus ($J^P =0^+$)
is written as $N_{\rm eff}^{\rm odd (1)}(\ell_{\pi}\otimes (J^P =0^+))$
as shown in Table~\ref{Odd_Even}.
For the even-N nuclear $^{122}$Sn target case,
the pionic atoms are populated in the nucleus with
a neutron-hole state $s_{1/2}^{-1}$,
and the effective number is written  
as $N_{\rm eff}^{\rm even}(\ell_{\pi}\otimes s_{1/2}^{-1})$.
By comparing the formula of these effective numbers,
we found the following relation,
\begin{equation}
N_{\rm eff}^{\rm even}(\ell_{\pi} \otimes s_{1/2}^{-1}) 
= 
2 N_{\rm eff}^{\rm odd (1)}(\ell_{\pi}\otimes (J^P =0^+)).
\label{s}
\end{equation}
The effective number of the even-N nuclear target 
is twice larger than that 
of the odd-N nuclear target.
This is simply because that
we assumed the number of neutron in the fully occupied 
$s_{1/2}$ orbit in $^{117}$Sn
is equal to be 1 
in the effective number formalism in Section~\ref{Formula} as explained in
Eq.~(\ref{initial}).
Therefore, in this case, 
we found that the cross section on the odd-N nuclear target 
is  half of that on the even-N nuclear target.

Then,
we consider the pionic atom formation
in the case  
of neutron pick-up from single particle orbits other
than $s_{1/2}$ in the target. 
For the odd-N nuclear $^{117}$Sn target case,
the pionic atoms are populated in
the nucleus with the configuration $J= [s_{1/2} \otimes j_h^{-1}]$,
where the neutron remaining in the valence $s_{1/2}$ state
couples to a neutron-hole state $j_h^{-1}$
to form the total nuclear spin $J$
which can take two values as
$J_{1} = j_h - \frac{1}{2}$ and $J_{2}=j_h + \frac{1}{2}$.
The effective numbers
for the pionic state ($\ell_{\pi}$) formation in
the daughter nucleus with the spin $J=J_1$, $J_2 $ are written
as
$N_{\rm eff}^{\rm odd(2)}(\ell_{\pi}\otimes [s_{1/2} \otimes j_h^{-1} ]_{J_1})$
and
$N_{\rm eff}^{\rm odd(2)}(\ell_{\pi}\otimes [s_{1/2} \otimes j_h^{-1}]_{J_2})$,
respectively.
For the even-N nuclear $^{122}$Sn target case,
the pionic atoms are populated in the nucleus with
a neutron-hole state $j_{h}^{-1}$
and the effective number is written
as $N_{\rm eff}^{\rm even}(\ell_{\pi}\otimes j_{h}^{-1})$.
In this case,
we found another relation written as,
\begin{equation}
N_{\rm eff}^{\rm even}(\ell_{\pi}\otimes j_{h}^{-1}) 
= 
N_{\rm eff}^{\rm odd(2)}(\ell_{\pi}\otimes [s_{1/2} \otimes j_h^{-1} ]_{J_1})
+ 
N_{\rm eff}^{\rm odd(2)}(\ell_{\pi}\otimes [s_{1/2} \otimes j_h^{-1} ]_{J_2}),
\label{jh}
\end{equation}
between the effective numbers for the pionic atom formation by 
the ($d,^3$He) reactions on the even-N and odd-N nuclear targets.
This relation means that the effective number of the even-N nuclear target
is equal to the sum of the effective numbers
of the odd-N nuclear target for the same combination of $\ell_\pi$ and 
$j_h^{-1}$.
Therefore, in this case, 
we found that the ($d,^3$He) total cross section obtained
by summing up the all subcomponents should take similar values both
for even-N and-N odd nuclear target cases.

Hence, as for the effective numbers,
we do not find the origin of the significant differences of
the absolute value of the ($d,^3$He) total cross sections
on the even-N and odd-N nuclear target cases
as shown in Eq.~(\ref{jh}),
except for the factor 2 difference of 
the contribution from the $s_{1/2}$ neutron
as shown in Eq.~(\ref{s}).

Finally, we consider the effects of
the additional factors $F_O$ and $F_R$
introduced in
Eqs.~(\ref{Neff_FoFr_1}) and~(\ref{Neff_FoFr_2}).  
To clarify the effects, 
we investigate the height of the largest peak structure appeared in the spectra
around $B.E. \simeq 3.9$ MeV in Fig.~\ref{Cross_0deg}
as an example and evaluate the effects of the $F_O$ and $F_R$ factors.
The largest peak mainly consists of the contributions of the neutron
pick-up from the $s_{1/2}$ state for 
both $^{122}$Sn and $^{117}$Sn target cases.
We pay special attention to the dominant subcomponent 
[$(1s)_\pi \otimes 0^{+}_{\rm ground}$] for $^{117}$Sn target
and the 
[$(1s)_\pi \otimes (3s_{1/2})^{-1}_n$] for $^{122}$Sn target.
According to Eq.~(\ref{s}), 
the effective numbers of these subcomponents satisfy the relation,
\begin{eqnarray}
\frac{N_{\rm eff}^{\rm odd(1)}([(1s)_\pi \otimes 0^{+}_{\rm ground}])}
{N_{\rm eff}^{\rm even}([(1s)_\pi \otimes (3s_{1/2})^{-1}_n]) }
=
\frac{1}{2} \ .
\label{s_ratio}
\end{eqnarray}
We consider the effects of the $F_O$ and $F_R$ factors to this ratio. 
From Eq.~(\ref{Neff_FoFr_1}),
the ratio is modified by including $F_O$ and $F_R$ as,
\begin{eqnarray}
\frac{N_{\rm eff}^{\rm odd(1)}([(1s)_\pi \otimes 0^{+}_{\rm ground}]) 
\times F_{O}^{\rm odd} \times F_{R}^{\rm odd}}
{N_{\rm eff}^{\rm even}([(1s)_\pi \otimes (3s_{1/2})^{-1}_n]) 
\times F_{O}^{\rm even} \times F_{R}^{\rm even}} 
=
\frac{1 \times 0.47 \times 0.68}{2 \times 0.7 \times 1}
\simeq \frac{1}{4},
\label{ratio}
\end{eqnarray}
where the $F_O$ and $F_R$ factors for $^{117}$Sn 
are $F_O^{\rm odd}=0.47$ and $F_R^{\rm odd}=0.68$ as listed 
in Tables~\ref{117Sn_Fo} and~\ref{117Sn_Fr},
and
for $^{122}$Sn they are $F_O^{\rm even}=0.7$ and $F_R^{\rm even}=1$ 
as reported in Tables V and VI of Ref.~\cite{Ikeno}.
%
By including $F_O$ and $F_R$, 
the height of 
the  largest peak of the spectra of the $^{117}$Sn target
is reduced to around 32\% ($= 0.47 \times 0.68$)
of its original value 
due to the effects of both factors. 
On the other hand,
as for the $^{122}$Sn target case,
the height is reduced to around 70\%
due to the effect of $F_O$.
The $F_R$ factor of $^{122}$Sn is equal to 1
and do not change the height of the peak.
Consequently,   
we found that the height of the largest peak structure 
of the $^{122}$Sn target
is expected to be about four times larger
than that of the $^{117}$Sn target by including the $F_O$ and $F_R$
factors as shown in Eq.~(\ref{ratio}).
We think this is the reason why the cross sections
in Fig.~\ref{Cross_0deg} show the differences between 
$^{122}$Sn and $^{117}$Sn targets.
We should notice here that the effect of $F_R$ is just to split one peak 
into several small peaks corresponding to the plural nuclear excitations
with the same quantum numbers.
Thus, the energy integrated total cross section is hardly changed by 
the $F_R$ factor.

We conclude that 
the origins of the difference of 
the absolute value of the cross sections
on the $^{117}$Sn and $^{122}$Sn targets
are the factor two difference of 
the effective numbers $N_{\rm eff}$ 
of the $s_{1/2}$ neutron pick-up process,
and the values of the $F_O$ and $F_R$ factors
which reflect the occupation probability
of the neutron states in the target and 
the relative strength of the excited levels of the daughter nucleus.
Here, we should be careful
on the differences of the effects of $F_O$ and $F_R$. 
The effect of $F_O$ makes the height of the peak lower
and also makes the energy integrated total cross section smaller,
whereas 
$F_R$ only divides the peak into small peaks and 
spreads in the energy spectrum. 
Hence, 
the value of the energy integrated total cross section 
changes little 
even if the $F_R$ factor is taken into account.

\end{document}